\documentclass[%
  aip,
  amsmath,amssymb,
  reprint,%
 superscriptaddress,
 longbibliography,
]{revtex4-1}

\usepackage{graphicx}
\usepackage{dcolumn}
\usepackage{bm}
\usepackage{hyperref}
\usepackage{xcolor}
\usepackage{soul,color}
\usepackage{verbatim}
\hypersetup{
    colorlinks=true,
    linkcolor=blue,
    citecolor=blue,
    urlcolor=blue,
}
\setcitestyle{numbers,square}

\begin{document}

\preprint{APS/123-QED}

\title[manuscript in preparation]{Optothermal control of spin Hall nano-oscillators}

\author{Shreyas~Muralidhar}
\affiliation{Department of Physics, University of Gothenburg, 412 96, Gothenburg, Sweden}

\author{Afshin~Houshang}
\affiliation{Department of Physics, University of Gothenburg, 412 96, Gothenburg, Sweden}

\author{Ademir~Alem\'an}
\affiliation{Department of Physics, University of Gothenburg, 412 96, Gothenburg, Sweden}

\author{Roman~Khymyn}
\affiliation{Department of Physics, University of Gothenburg, 412 96, Gothenburg, Sweden}

\author{Ahmad~A.~Awad}
\email{ahmad.awad@physics.gu.se}
\affiliation{Department of Physics, University of Gothenburg, 412 96, Gothenburg, Sweden}

\author{Johan~\AA{}kerman}
\affiliation
{Department of Physics, University of Gothenburg, 412 96, Gothenburg, Sweden}%

\date{\today}

\begin{abstract}

We investigate the impact of localized laser heating on the auto-oscillation properties of a 170 nm wide nano-constriction spin Hall nano-oscillators (SHNO) fabricated from a NiFe/Pt bilayer on a sapphire substrate. A 532 nm continuous wave laser is focused down to a spot size of about 500 nm at a power ranging from 0 to 12 mW. Through a comparison with resistive heating, we estimate a local temperature rise of about 8 K/mW. We demonstrate reversible laser tuning of the threshold current, the frequency, and the peak power, and find that the SHNO frequency can be tuned by up to 350 MHz, which is over three times more than the current tuning alone. Increasing the temperature also results in increased signal jitter, an increased threshold current, and a reduced \emph{maximum} current for auto-oscillations. Our results open up for optical control of single SHNOs in larger SHNO networks without the need for additional voltage gates. 

\end{abstract}

\pacs{Valid PACS appear here}
\keywords{Optical control, tuning, spin Hall effect, nano-oscillators, SHNO}

\maketitle

Advances in spintronics have led to the development of spin current driven, energy efficient, and nano-scale microwave oscillators \cite{Berger1996,Slonczewski1996,Slonzewski1999jmmm,Tsoi1998,Myers1999,Kiselev2003,Rippard2004prl,Katine2008,Demidov2012b, Dumas2014,Silva2008,chen2016ieeerev,Duan2014natcomm,Haidar2019natcomm,fulara2019sciadv,shao2021ieeetmag}, with applications \cite{Dieny2020ntelec} in diverse domains such as magnetic memory, ultra-wide band microwave signal generation \cite{bonetti2009, bonetti2010prl} and detection \cite{Tulapurkar2005nt,albertsson2020ieeetnano}, and most recently in neuromorphic computing \cite{Torrejon2017Nature,Romera2018nt,Grollier2016procieee,Zahedi2020Ntn,chumak2021roadmap} and Ising machines \cite{Albertsson2021apl,houshang2020spin,Zahedinejad2022natmat} using linear and two-dimensional oscillator arrays. Nano-constriction based spin Hall nano-oscillators (SHNOs) \cite{Demidov2014,Zahedinejad2018apl,hache2019apl,sato2019prl, hache2020apl} are one such class of easily tunable \cite{Zahedi2017ieee,goncalves2021prappl} microwave signal generators, which exhibit particular advantages such as simple fabrication, good thermal dissipation, robust mutual synchronization \cite{Awad2016}, and easy direct access to the magnetodynamically active region, \emph{e.g.}~for voltage gating and/or optical access.  

Whereas voltage gating has recently been demonstrated as a way to control and program individual nano-constrictions in SHNO arrays \cite{liu2017prappl,Fulara2020_natCom,Zahedinejad2022natmat}, one may also envision optical control as an alternative means to program one or more of the constituent oscillators, \emph{e.g.}~through direct heating. To this end, a better understanding of the impact of temperature on the SHNO auto-oscillation frequency, linewidth, and power is key to any such implementation. 
Early experimental studies on temperature-induced effects in STNOs were reported in \cite{Sankey2005prb,Mistral2006apl} and detailed theoretical analysis of thermal noise was performed in \cite{Jiang2005prb,Russek2005prb,Kim2008prl,Tiberkevich2007c} where it was shown that the linewidth increases with an increase in thermal noise. While similar arguments should hold for SHNOs, any experimental studies have not yet been reported.  

Here, we present a detailed study of the effects of local laser heating of a nano-constriction SHNO. By focusing a laser down to about a 500 nm spot size on top of the nano-constriction, and varying the laser power from 0 to 12 mW, we are able to tune the frequency by up to 350 MHz. In addition, laser heating is found to increase the threshold current and to limit the current range for auto-oscillations also from the high current end. Our results provide a route towards optical control of much larger ensembles of SHNOs, where instead of a single laser spot, an entire image with different optical intensity can be projected onto a large SHNO array for parallel control of all individual nano-constrictions without any need for voltage gates. 

\begin{figure}
    \centering
\includegraphics[width=\linewidth]{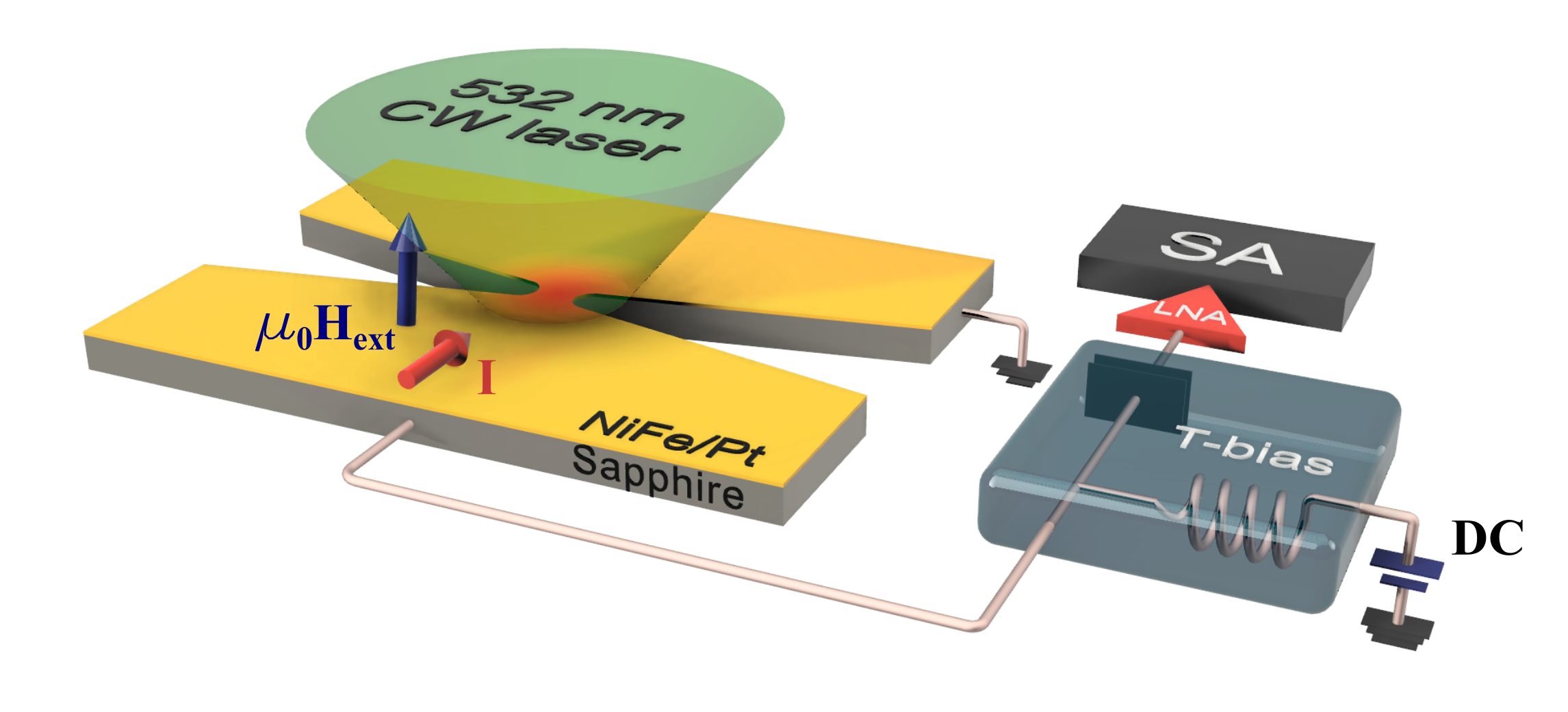}
\caption{\label{fig:intro} Schematic of the experimental setup where the nano-constriciton SHNO is locally heated using a focused green laser. A direct current, injected through a bias-T, drives auto-oscillations of the NiFe layer, which through the AMR generates a microwave voltage that is amplified and then recorded using a spectrum analyzer. }
\end{figure}

SHNO stacks of NiFe(5nm)/Pt(5nm) were prepared on sapphire substrates using dc magnetron sputtering under a 3~mTorr Ar atmosphere in an ultra-high vacuum chamber with $1.5\times 10^{-8}$~mTorr base pressure. Single 170 nm wide nano-constrictions were defined using electron beam lithography, and the pattern was transferred to the stack using ion beam etching \cite{Kumar2022nanoscale}. Finally, a conventional ground--signal--ground (GSG) waveguide and electrical contact pads for wide frequency range microwave measurement were fabricated by photolithography, Cu/Au deposition, and lift-off.

A schematic of the microwave measurements is shown in Fig.1. The SHNO drive current is injected through a bias-T and the microwave signal from the resulting auto-oscillations is amplified by a +70 dB LNA and then recorded using a spectrum analyzer. The sample holder contains a ceramic heating element and is mounted on a 3-axis motorized precision positioning stage. A 532 nm continuous wave (CW) laser is focused on the sample through a high numerical aperture (NA = 0.75) objective resulting in a spot diameter of $\sim$500 nm. The laser power is controlled by a polarizer and a half wave plate mounted on a motorized rotational mount. The optics are carefully aligned and the movement of the motor is automated and calibrated to extract the laser power reaching the sample.

The local temperature at the SHNOs during laser heating is obtained by performing a calibration, with the laser off, of the auto-oscillation frequency \emph{vs.}~temperature up to 373 K, using the ceramic heater. The temperature is stabilized using a Lakeshore 335 controller and a PT100 temperature sensor.  

\begin{figure}[b]
    \centering
\includegraphics[width=0.8\linewidth]{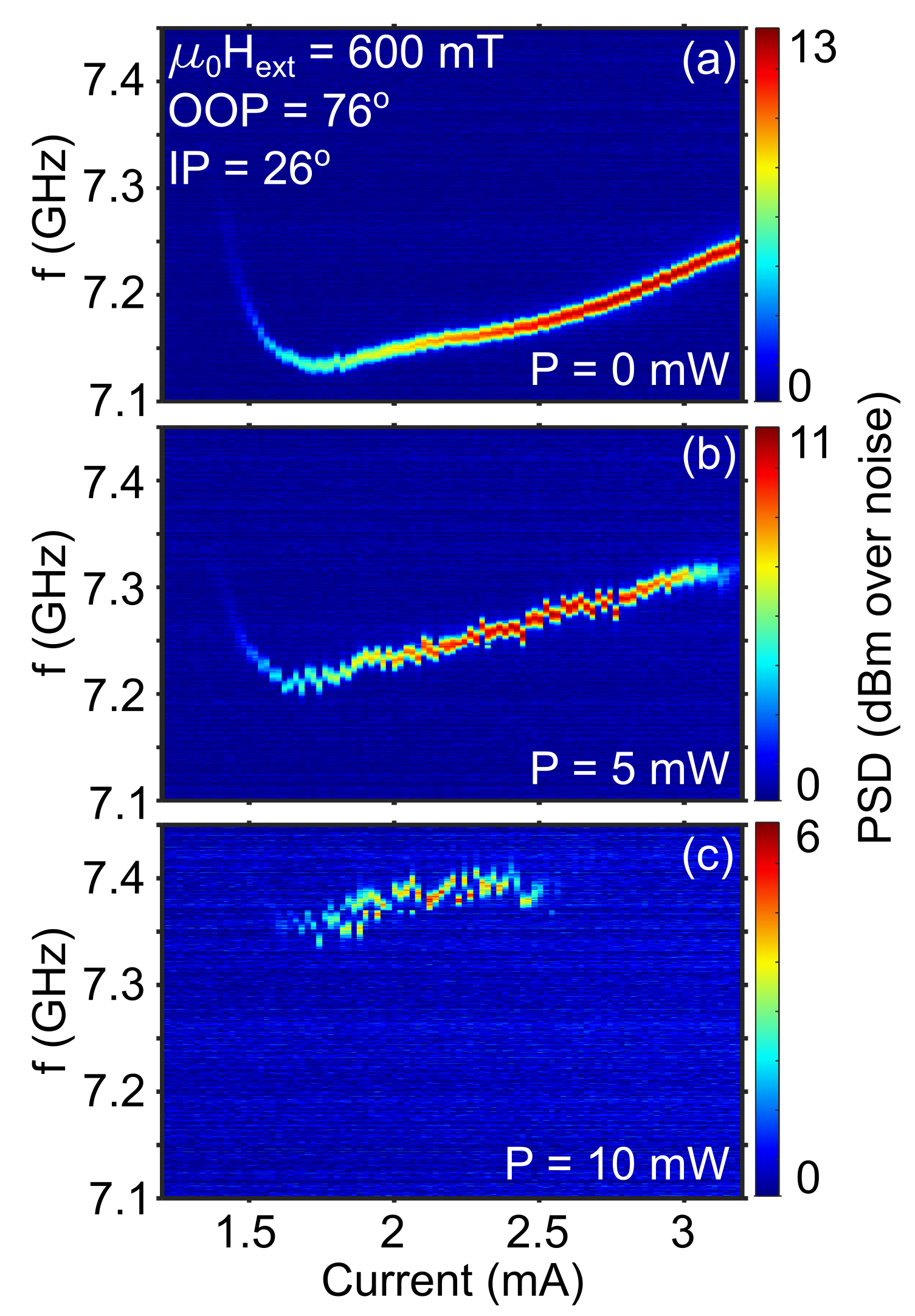}
\caption{\label{fig:PSDvsI} Power spectral density (PSD) \emph{vs.}~current at three different laser powers: (a) 0 mW, (b) 5 mW,  and (c) 10 mW. An oblique magnetic field of 600 mT is applied at a 76\textdegree \space out-of-plane (OOP) angle and an in-plane (IP) angle of 26\textdegree.}
\end{figure}

The power spectral density (PSD) of the SHNO auto-oscillation, without any laser heating, is shown \emph{vs.}~current in Fig.2(a), for a magnetic field of 600 mT, applied at an out-of-plane (OOP) angle of 76$^\circ$ and an in-plane (IP) angle of 26$^\circ$, measured relative to a direction perpendicular to the current. The expected non-monotonic frequency behavior is observed, with a red-shifting frequency up to 1.7 mA, after which the frequency increases quasi-linearly, consistent with an edge mode moving into the nano-constriction interior and then further expanding \cite{Awad2020APL,Dvornik2018prappl}. 

\begin{figure}[b]
    \centering
\includegraphics[width=0.8\linewidth]{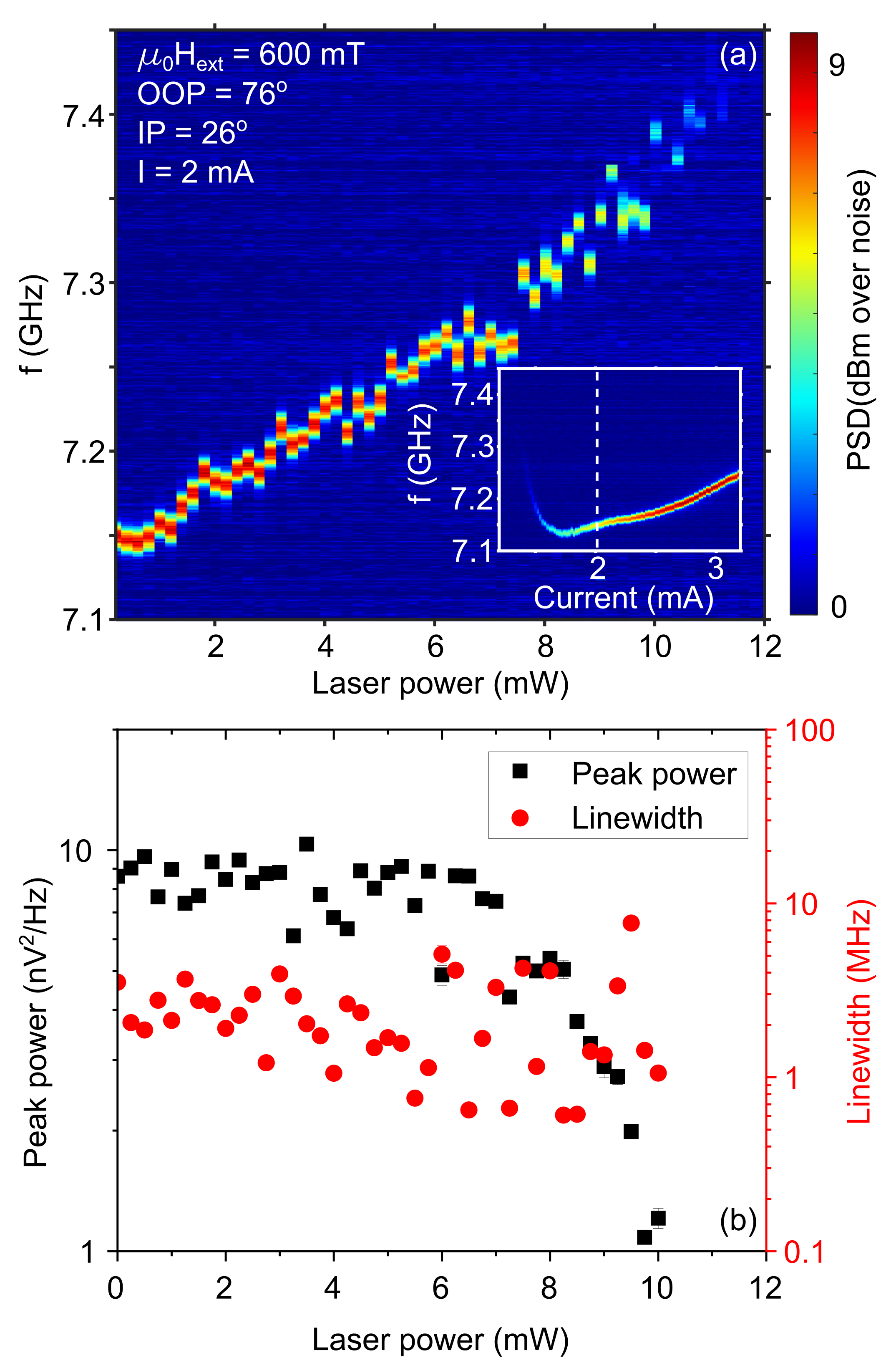}
\caption{\label{fig:PSDvsPower} (a) PSD  \emph{vs.}~laser power at a constant current of 2 mA. The white dashed line in the inset shows the current operating point. (b) Extracted peak power and linewidth as a function of  laser power. }
\end{figure}

Fig.2(b) shows the same device when illuminated with 5 mW. While the overall shape of the frequency behavior stays the same, the entire curve is shifted $\sim$80 MHz to higher frequencies and there is pronounced frequency jitter indicating an increased $1/f$ noise. We also note that the auto-oscillations now gradually disappear at the high current end, \emph{i.e.}~in addition to a threshold current, there also exists a maximum current beyond which auto-oscillations are no longer observed. In Fig.2(c) we have increased the laser power to 10 mW, which shifts the curve another $\sim$150 MHz, increases the frequency jitter considerably, and greatly reduces the current range for auto-oscillations from both ends.

Figure.~\ref{fig:PSDvsPower}a shows a plot of the auto-oscillation PSD \emph{vs.}~laser power at a single operating point of 2 mA in the same magnetic field as in Fig.2; the inset shows the original PSD \emph{vs.}~current without any laser heating. We have intentionally used the same frequency scales in the main figure and the inset to clearly show the greater frequency tunability from laser illumination compared to using only the current. While the current can tune the auto-oscillation frequency $\sim$120 MHz, laser powers of up to 11 mW can achieve $\sim$250 MHz of frequency tuning, above which the auto-oscillations disappear. As already indicated in Fig.2, the frequency jitter increases gradually with laser power. We extract both the auto-oscillation linewidth and the peak power from fits of a single Lorentzian to the PSD and show the results on log scales in Fig.3(b). While the variation in the extracted linewidth seems to increase continuously from zero laser power, the average linewidth remains largely constant. The peak power also remains largely constant up until 7 mW, above which it shows a rapid exponential decay with current. 

\begin{figure}[t]
    \centering
\includegraphics[width=0.95\linewidth]{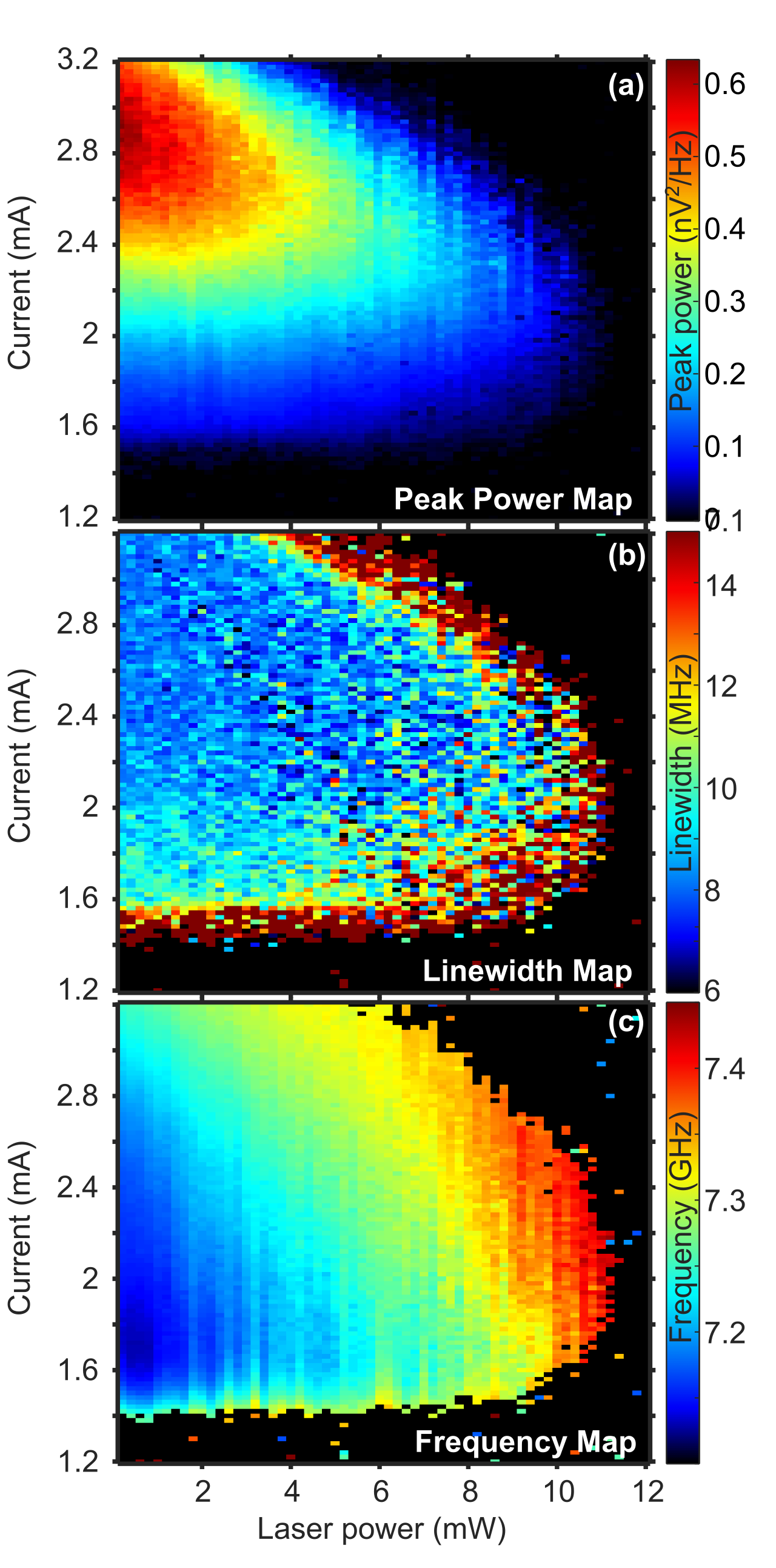}
\caption{\label{fig:Maps} Simultaneous current and laser tuning of the SHNO with extracted auto-oscillation (a) peak power, (b) linewidth, and (c) frequency.}
\end{figure}

It is then interesting to map out the combined effect of current tuning and external laser heating. Fig.4 shows 2D color maps of the frequency, linewidth, and peak power $\emph{vs.}$~both current and laser power. From the map of the peak power in Fig.4(a) (here on a linear scale), and in particular from its contour lines of equipower, we can more clearly see how the laser affects both the threshold current and the maximum current for auto-oscillations. We also note that the qualitative impact from current and laser tuning appears otherwise to be rather similar. 

From the corresponding 2D map in Fig.4(b) of the linewidth we instead note that the laser has a more direct impact on the linewidth \emph{variation} compared to the current. This correlates with the increasing frequency jitter and it appears as if local laser heating has a much more detrimental effect on this jitter than the current. The average linewidth is otherwise relatively constant in the interior of the map, shows a greater tunability with the current than with the laser power, and finally increases rapidly towards the boundary of auto-oscillations, correlating well with the contours of low peak power in Fig.4(a). While the increased frequency jitter might be viewed as a drawback, it is an interesting mechanism for injecting noise to an invidual SHNO within an array. It might also open up for novel optical annealing schemes applicable to Ising machines. Simultaneous laser heating of all SHNOs in an Ising machine would then not affect their relative frequency differences, and hence not their couplings, but increase the effective annealing temperature through the increased frequency jitter.

In Fig.4(c), the much more effective tuning of the frequency with laser heating is again directly observed. However, through the combined tuning, using both current and laser illumination, we can extend the total frequency tuning range to 350 MHz: The lowest auto-oscillation frequency of 7.1 GHz is observed at 1.8 mA and 0 mW, while the highest frequency of 7.45 GHz is observed at 2.4 mA and 10.5 mW. The maximum laser tuning range is observed for currents between 1.7 and 2.5 mA. 

\begin{figure}[t]
    \centering
\includegraphics[width=\linewidth]{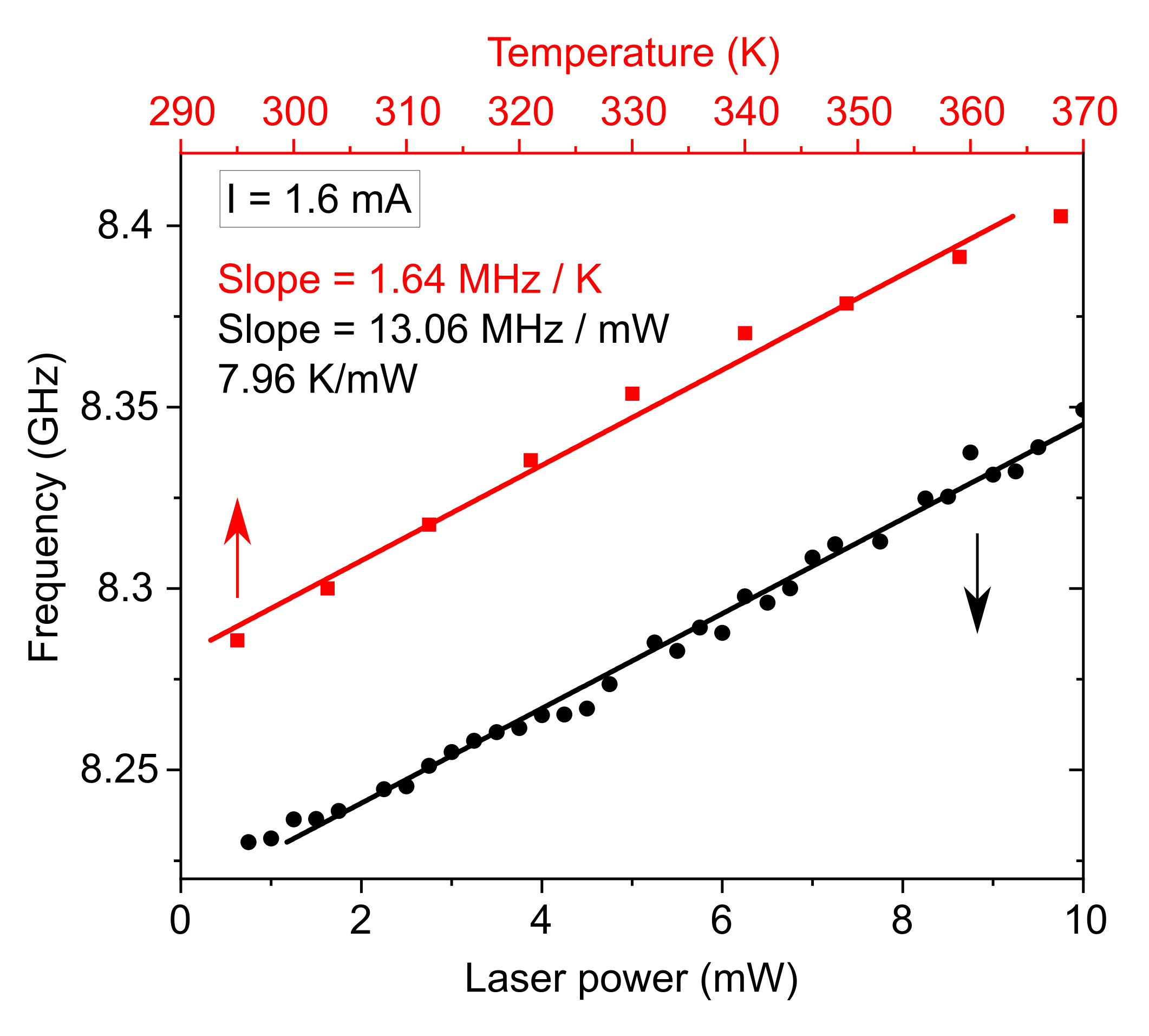}
\caption{\label{fig:freqvsPowerT} Auto-oscillation frequency as a function of  laser power (in solid black circles) and as function of temperature (in solid red squares). Solid black and red lines are the linear fits for these data respectively.}
\end{figure}

Finally, to estimate the temperature rise from the laser heating, we compared the auto-oscillating frequency under laser heating with that measured using resistive heating. Any laser induced currents can be safely ignored as the photoelectric effect of a sub-micron diameter linearly-polarized laser spot on such nanoscale samples fabricated on sapphire should be negligible. 

Figure \ref{fig:freqvsPowerT} shows the frequency shift due to laser heating, depicted by the solid black circles, compared to resistive heating shown by the solid red squares; the corresponding $x$ axes are shown in black and red respectively. From the linear fits to both data sets we can conclude that every mW of laser power heats the nano-constriction by about 8 K. 

In conclusion, we have demonstrated contactless optical tuning of the auto-oscillation signal properties of a single nano-constriction SHNO. Using only laser tuning, we can vary the SHNO frequency by 250 MHz. Through a combination of current and laser tuning, the maximum tuning range reaches 350 MHz, which is more than three times that of current tuning alone. The laser heating additionally tunes the operational current range of the SHNO, increasing the threshold current and limiting the current range for auto-oscillation at high currents. Laser heating is found to increase the 1/f noise of the SHNO more than current tuning does. While this might be viewed as a drawback, it also opens up for novel optical annealing schemes applicable to Ising machines. Laser tuning of SHNOs hence provides an additional degree of freedom for controlling a single SHNO in large arrays of SHNOs, where temporal and/or spatial intensity modulation can be projected onto the array for temporal and parallel control of all individual nano-constrictions without any need for physical gate contacts.

\section*{Acknowledgments}
This work was partially supported by the Horizon 2020 research and innovation programmes No. 835068 "TOPSPIN" and No.~899559 "SpinAge". This work was also partially supported by the Swedish Research Council (VR Grant No. 2016-05980) and the Knut and Alice Wallenberg Foundation. 

\section*{Data availability}
The data that support the findings of this study are available from the corresponding author upon request.
%

\end{document}